\documentclass[pre,aps,twocolumn,showpacs,floatfix]{revtex4}
\usepackage{epsfig}
\usepackage{graphicx}
\usepackage{amsmath}

\begin{document}

\title{Ab initio study of the atomic motion in liquid metal surfaces: comparison with Lennard-Jones systems.} 

\author{Luis E. Gonz\'alez and David J. Gonz\'alez} 
\affiliation{Departamento de F\'\i sica Te\'orica, 
Universidad de Valladolid, 47011 Valladolid, SPAIN.}
\date{\today}

\begin{abstract}
It is established that liquid metals exhibit surface layering
at the liquid-vapor interface, while dielectric simple systems, like those interacting through Lennard-Jones potentials, show a monotonic decay from the liquid density to that of the vapor.
First principles molecular dynamics simulations 
of the liquid-vapor interface of several liquid metals (Li, Na, K, Rb, Cs, Mg, Ba, Al, Tl and Si), and the Na$_3$K$_7$ alloy
near their triple points have been performed in order to study the 
atomic motion at the interface, mainly at the
outer layer. Comparison with results of classical molecular dynamics simulations of a Lennard-Jones system shows interesting differences and similarities.
The probability distribution function of the time of residence in a layer 
shows a peak at very short times and a long lasting tail. The mean residence time in a layer increases when approaching the interfacial region, slightly in the Lennard-Jones system, but strongly in the metallic systems. The motion 
within the layers, parallel to the interface, can be described as 
diffusion enhanced (strongly, in the case of the outermost layer) with respect to 
the bulk, for both types of systems, despite its reduced dimensionality in metals.
\end{abstract}


\maketitle

\section{Introduction.}

Surface layering at metallic liquid-vapor interfaces was
suggested in the early eighties by Monte Carlo simulations of alkali metal 
liquid surfaces \cite{develyn:prl47:44}.
Subsequently, more elaborate Monte Carlo techniques (see \cite{rice:ms29:93}
and references therein), experimental measurements 
\cite{magnussenetal:prl74:44,reganetal:prl75:98,
reganetal:prb55:74,tostmannetal:prb59:83,
tostmannetal:prb61:84,dimasietal:prl86:38,
shpyrkoetal:prb67:05,shpyrkoetal:prb70:06}
and recent ab initio molecular dynamics (MD) simulations
\cite{fabriciusetal:prb60:83,walker:jpcm16:75,
gonzalez:prl92:01,gonzalez:prl94:01,gonzalez:jcp123:01}, 
have established that the ionic density profile of liquid metals 
across the interface shows oscillations that decay into the bulk
liquid after, on average, three to four layers.
The use of ab initio MD simulations in these studies is 
important for two reasons. First, being ab initio, the valence 
electrons and the ions are treated  
on the same footing, with the ions reacting 
consistently to the large spatial variations undergone by the electronic 
density when 
moving from the liquid to the vapor. 
Second, being MD simulations, a study of the motion of the atoms 
can be carried out directly and changes across the interface
can be analyzed. This type of study has not yet been undertaken, as far as we know, basically because most of the simulations of liquid metallic surfaces were carried out using the Monte Carlo method, which gives no information about the dynamic properties of the system studied.
The main aim of this paper is to fill this gap, reporting a study of the atomic motion in the liquid-vapor interfaces of Li, Na, K, Rb, Cs, Mg, Ba, Al, Tl, Si, and 
the Na$_3$K$_7$ alloy, using slabs with 2000 atoms.

Layering also appears for metals at the solid-liquid
interface \cite{jesson:jcp113:35}, for liquids in contact with a hard wall 
\cite{huismanetal:nat390:79,yuetal:prl82:26}, or in confined geometries \cite{teng:prl90:04}.
Moreover, for ultrathin films of large molecules in contact with a Si wall
\cite{schuster:europolymerj40:93}
and for confined dusty plasma liquids \cite{teng:prl90:04} it has 
been possible to follow experimentally the molecular motion, whereas in the solid-liquid 
interface ab initio MD simulations allowed to study the atomic diffusion 
at the interface \cite{jesson:jcp113:35}. In all cases reduced 
diffusion with respect to the bulk was found.

These oscillating profiles are in contrast with those of the 
liquid-vapor interfaces of one-component dielectrics, like water or Lennard-Jones (LJ) systems, 
or those of liquid-liquid interfaces of immiscible mixtures, which show monotonic profiles 
with no layering.
Taking the archetypical example of the one-component LJ system, we have found in the literature a large amount of computer simulation studies of the coexisting densities, the interfacial width and the surface tension, as a function of the temperature, the particular LJ model used (truncated, truncated and shifted, full potential) and the lateral area simulated (see, for instance, \cite{gloor:jcp123:03} and references therein).
However, to our knowledge, an analysis of the atomic motion in the interface has not been performed yet.
Surprisingly, the situation is somewhat different for more complex systems, like water,
or mixtures, where some (scarce) studies of atomic diffusion in the interfacial region have indeed been performed \cite{meyer:jcp89:67,buhn:flpheq224:21,benjamin:jcp97:32,liu:jpcb108:95}.
In the case of mixtures, some anisotropy in diffusion was detected, either due to an increase in the diffusion coefficient parallel to the interface, $D_T$, with respect to the bulk value \cite{meyer:jcp89:67,buhn:flpheq224:21}, or to a decrease of the diffusion coefficient normal to the interface, $D_N$, \cite{benjamin:jcp97:32}.
Following a thorough analysis of the effect of the inhomogeneity of the interface on the values (and even the definition) of $D_N$, molecular dynamics simulations of the liquid-vapor interface of water \cite{liu:jpcb108:95} revealed an increase in $D_T$ (3.5-fold) and also in $D_N$ (2-fold), which was attributed to the reduction of the number of hydrogen bonds in the interface. Moreover, it was argued that the same qualitative behavior would occur for other simple liquids \cite{liu:jpcb108:95}. 

Although surface layering is common to the liquid-vapor interface of metals and
the systems mentioned above, there are also important differences. 
Even though the atoms in the outer layer of the liquid-vapor interface rarely 
leave the liquid surface (especially at temperatures near their triple points),
they are not geometrically confined. Moreover,
in liquids confined, or on a wall, or at the interface with their solid phase, a strong interaction between the substrate and 
the first layers of the liquid has an important influence on the structure and 
dynamics of the liquid, leading to strong oscillations in the profile and to
reduced diffusion. 
It is also worth analyzing if the
metallic liquid-vapor interface, in particular the outer layer, behaves 
like a quasi two-dimensional system. There are some liquid 
alloys (Ga with small
amounts of Tl, Pb or Bi) where the minority component, which has
a high melting temperature and segregates to the surface,  displays this 
kind of behavior 
(see, e. g., references 
\cite{yang:prb67:03,issanin:cpl394:20,issanin:jcp121:05}).
For one component metals or other type of alloys, a possible 
measure of the two-dimensional character of the outer layer could be some 
distinctive feature of the density distribution function (DDF) of the time 
of residence of the atoms in that layer, for
instance its mean value.

\section{Simulation details.}

Because of the the periodic boundary conditions used in most simulation 
methods, a slab geometry is usually adopted in studies of the liquid-vapor 
interface, with two free surfaces perpendicular to one of the axes, 
taken here as the $z$ axis. 
These two surfaces should be well separated to minimize interaction between them.
For the metallic systems we have performed ab initio molecular dynamics simulations, whereas for the LJ system standard classical molecular dynamics have been used.

Ab initio simulations based on the Kohn-Sham formulation of
Density Functional Theory (DFT) \cite{kohn:pr140:33}, 
pose huge computational demands, and the only
metallic liquid surfaces studied so far, Si \cite{fabriciusetal:prb60:83}
and Na \cite{walker:jpcm16:75},  
used small samples (96 and 160 particles respectively) where 
the two surfaces are rather close (16 and 20 \AA\ respectively). 
Orbital free ab initio molecular dynamics (OFAIMD) simulations 
reduce somewhat these computational demands, by returning to the 
original Hohenberg-Kohn formulation of DFT \cite{hohenberg:pr136:64}, 
and adopting an explicit but approximate density functional for the 
electron kinetic energy so that the whole energy is a functional of the 
electron density. OFAIMD simulations of liquid metal surfaces have recently been  
performed \cite{gonzalez:prl92:01,gonzalez:prl94:01,gonzalez:jcp123:01}
for Li, Na, Mg, Al, Si and Na$_{3}$K$_{7}$ using 2000 particles and for 
Li$_{4}$Na$_{6}$ using 3000 particles, which led to simulation boxes 
big enough to reliably represent a macroscopic interface. Details about the formalism and the electron-ion interactions can be found elsewhere \cite{gonzalez:jpcm13:01,gonzalez:prb65:01}.
Even though the OFAIMD method has been applied successfully to bulk metals and alloys, it might be argued that this is not a sufficient validation of the method, as surfaces are different from bulk systems. However, it must be stressed that the OFAIMD studies of Na and Si, produced results very similar to those obtained by the (in principle) more accurate Kohn-Sham {\em ab initio} simulations: the wavelength of the oscillations in the profiles, which is recovered exactly, the number of nearest neighbors of a Si particle across the interface, and the surface tension of liquid Na, are all well reproduced by the OFAIMD approach. Further confidence in the ability of the methd to tackle metallic surfaces can be obtained from studies for finite systems, where the surface plays an essential role. For instance, a long standing and previously unexplained anomalous variation of the melting temperatures of Na clusters with size \cite{Haberlandetal} has been reproduced and rationalized for the first time  in terms of the surface geometry and stability \cite{Aguado} using the same OFAIMD method as in this paper. Even furher confidence in the capabilities of
the method, concerning specifically semiintinite surfaces, can be gained from
preliminary results \cite{AlMg} obtained for the temperature dependent 
surface relaxation of Al($110$) and Mg($10\bar{1}0$), which reproduce 
qualitatively both experimental data and previous Kohn-Sham calculations. 

The LJ system has been simulated at a reduced temperature $T^*=k_BT/\epsilon=0.73$,
which is near its triple point, in order to compare with the metallic systems in a
similar thermodynamic state. We have made simulations with 1960 particles, similar to the metals, and also with 15680 particles in order to reassess the results for the smaller system. The lateral side for the small system is $11.75\sigma$, which is already large enough to suppress the unrealistic oscillatory behaviour of the interfacial properties due to periodic  boundary conditions \cite{Oreaetal}. For the large system the lateral side is doubled ($23.5\sigma$) so the expected effects on the interfacial properties are only those of the enhanced capillary waves which should produce a wider interface than in the small system.

For all systems, the number of $NVE$ equilibrium configurations used for averaging ranged between 20000 and 30000.


\section{Layers definition.}

\begin{figure}
\begin{center}
\mbox{\psfig{file=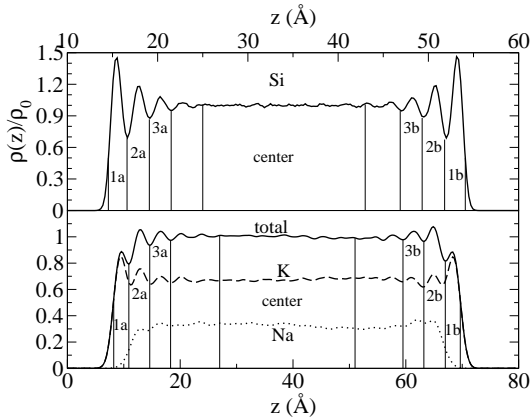,angle=-90,width=85mm}}
\end{center}
\caption{Ionic density profiles normal to the liquid-vapor 
interfaces of Si and Na$_3$K$_7$. 
The different regions where the atomic motion has been studied are also shown.} 
\label{layers}
\end{figure}

\begin{figure}[h]
\begin{center}
\mbox{\psfig{file=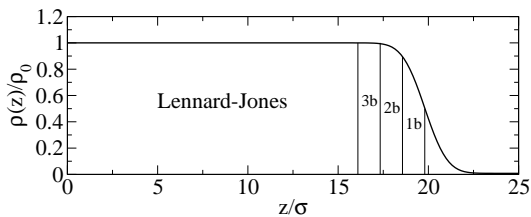,angle=-90,width=85mm,clip}}
\end{center}
\caption{Ionic density profile normal to the liquid-vapor 
interfaces of the LJ system at $T^*=0.73$, and the ``layers" defined for this interface. 
Only half the slab is shown.} 
\label{LJ}
\end{figure}

Figure \ref{layers} shows the ionic density profile 
obtained for liquid Si and the partial
and total ionic density profiles for the liquid alloy Na$_3$K$_7$, while that of the LJ system is depicted in figure \ref{LJ}.
The layers where the 
ionic motion was studied, which for metals were located between 
consecutive minima in the total ionic density profile, are also indicated. The outermost layer (numbered 1) comprises from the last minimum to the inflection point of the decaying profile.
For layers 1, 2 and 3 the results of the two layers on opposite sides of the slab
were averaged. Further structural details of the interfaces of these systems 
will be given elsewhere \cite{prbinpress}, 
but we note here that the relative amplitudes of the
outermost oscillation for the alkalis and alkaline-earths are rather similar 
(maxima from 1.00 to 1.13, minima from 0.80 to 0.88) while they increase 
significantly for the systems of valence 3 and 4 (maxima 1.29, 1.47 and 1.56, 
minima 0.73, 0.69 and 0.54 for Al, Si and Tl respectively).

\begin{figure}
\begin{center}
\mbox{\psfig{file=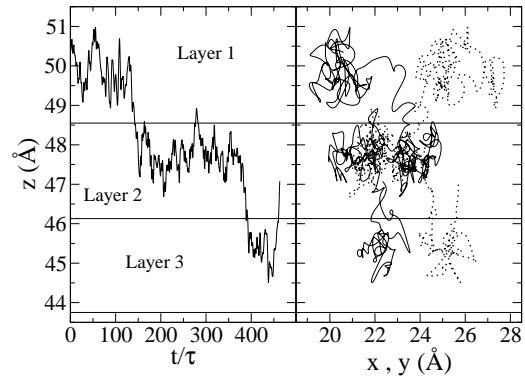,angle=-90,width=85mm}}
\end{center}
\caption{Time evolution of the $z$ coordinate of an Al atom and its projected
trajectory onto the $xz$ (full line) and the $yz$ (dotted line) planes.} 
\label{movim}
\end{figure}

The width of the different layers is very small, of the order of
one atomic diameter. Therefore it is not appropriate to talk about
diffusion along the direction normal to the interface within a layer, as  
there is no room for a particle within the layer to reach that type of  
motion. Figure \ref{movim} shows the time evolution of the $z$ coordinate 
of one particular Al atom and an oscillatory-type motion within the layers 
can be seen followed by jumps into adjacent layers. 
The trajectory of the particle, projected onto 
the $xz$ and $yz$ planes is also shown in the figure.
To provide a meaningful comparison among the results for different 
systems, with diverse 
masses and potentials, and consequently different characteristic times, 
a reference time, $\tau$, has been defined for each system which is to 
be used as a time unit (see table \ref{tablilla}). 
Specifically, $\tau$ has been taken as the first value of $t$ for 
which the mean square displacement (MSD) in the bulk liquid has an 
inflection point,
which is an indication of the atomic motion changing from free-particle 
to diffusive like behavior. 

The ionic density profile for the large LJ system is shown in figure \ref{LJ}.
For this type of system, the definition of ``layers" in order to analyze atomic motion is basically arbitrary. We have chosen to define them in a similar way as we have done for metals in order to make the comparison as fair as possible. First we have fitted the simulation results to an error function profile, 
$$\rho(z)=\frac12(\rho_{\ell}+\rho_v)+\frac12(\rho_{\ell}-\rho_v) \mbox{erf}(\sqrt{\pi} (z_0-z)/w) $$
(where $\rho_{\ell}$ and $\rho_v$ are the coexisting densities of the liquid and the vapor, $z_0$ is the position of the interface, which coincides with the inflection point of the profile, and $w$ its width),
which appears to be more adequate than the usual hyperbolic tangent one \cite{Sidesetal}. The widths for the small and large LJ systems are $w=2.25\sigma$ and $w=2.47\sigma$ respectively, the latter being larger as expected from the 
increased number of capillary waves. In the following we will present the results obtained for the large system, which, appart from this interfacial width, are practically coincident with those of the small system.
Layer number 1 is then defined as a slice of width $w/2$ from the inflection point towards the liquid. Further slices of width $w/2$ towards the liquid are then taken as layers 2 and 3.
Note that, defined this way, the layers of the LJ system are also very thin ($1.235\sigma$), so we are in a similar situation as in metals, where we consider that diffusion is only possible within the layers parallel to the interface. The reference time $\tau$ for the LJ system is finally defined in the same way as for metals, and is also shown in
table \ref{tablilla}.

\section{Results and discussion.}

In order to analyze the diffusive motion within the layers, the MSD along the 
$x$ and $y$ directions has been calculated for those particles that remain inside
the layer for a large enough time ($t_m$), a criterion somewhat different from, 
although in the same spirit as others used in previous works 
\cite{benjamin:jcp97:32,liu:jpcb108:95}. 
The selection of $t_m$ is somewhat delicate, as 
$t_m$ must be large enough so that diffusive motion has already set in, but the 
number of particles that remain in a layer for at least $t_m$ decreases as 
$t_m$ increases, leading to poorer statistics. A good compromise was found to
be $t_m=t_{20}$, defined so that the probability of residence in the layer for 
times larger than $t_{20}$ is 20 \%.
All the data shown below were obtained with this criterion, but checks showed 
that other reasonable choices for $t_m$ led to statistically equivalent 
results. In order to compute $t_{20}$, and also to characterize the motion along
the $z$-direction, the residence time of particles in each of the layers was 
studied. 

\begin{figure}
\begin{center}
\mbox{\psfig{file=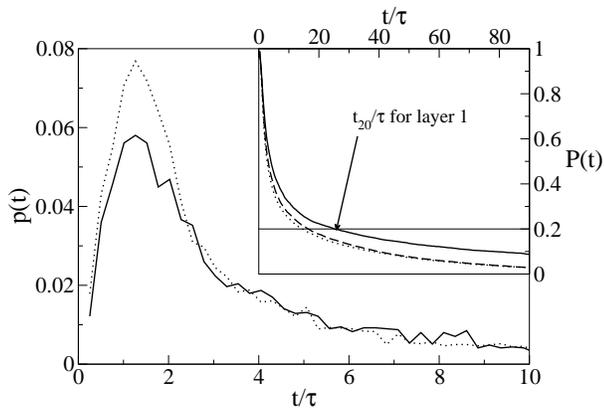,angle=-90,width=85mm}}
\end{center}
\caption{Distribution function of the residence time, $p(t)$, 
in layer 1 (full line) and layer 3 (dotted line) for liquid Ba. 
The inset shows the integrated probability 
distribution function, $P(t)$, together
with the 20 \% level used to define $t_{20}$. The dashed line denotes
layer 2.} 
\label{probab}
\end{figure}

\subsection{Motion perpendicular to the interface.}

For all the systems considered, including the metals and the LJ system, 
the DDF of the residence time, $p(t)$, has been computed  for the three outer layers (the case of Ba is shown in figure \ref{probab}), and some common features can be identified. 
The maximum of $p(t)$ occurs at very small $t$-values, ($\approx$ $\tau$), 
suggesting that most of the particles attempting to enter the layer are bounced back. Also, the DDF  exhibits a long time tail, which gives rise to mean 
residence times (MRT), $t_{\rm av}$, which increase when moving from the 
bulk liquid towards the outer layers (see table \ref{tablilla}). 
In the LJ system this increase is rather small, amounting to approximately a 10 \%.
For the metals, however, the MRT at the outermost layer is strongly enhanced, with increases between 75 \% (for Na) and 246 \% (for Tl), while the values taken at layer 2 show a more modest increase of around 10 \%.
This large discrepancy between the behavior of the LJ system and that of metals precludes any simple geometrical interpretation of the results and underlines that the change of interactions at the liquid-vapor interface of metals does have an important influence on some dynamic properties of the surface.
Moreover, these results indicate that the outermost layer in metals behaves more bidimensional-like than the inner ones. 
It is interesting to note that the MRT is around 20$\tau$ for the alkalis 
and alkaline-earths, whereas it increases in Al ($\approx 40\tau$) and Tl 
($\approx 60\tau$), but not in Si. 

In order to rationalize these differences among metals, we consider how the atoms
move from one layer to another. The minima in the density profiles suggest 
the existence of an interlayer potential barrier, which would be higher the 
lower the minimum of the density profile. Therefore this barrier would 
increase significantly from the alkalis and alkaline-earths to Al, to Si, and 
finally to Tl. The probability of overcoming this barrier would be related 
both to its height and to the frequency with which the atoms attempt to cross. 
This frequency has been estimated as proportional to the Einstein frequency along 
the $z$ direction, $\Omega_z$, which is shown in table \ref{tablilla} together with 
the corresponding frequency in the direction parallel to the interface, 
$\Omega_T$, for completeness. These have been obtained from a short time (up to $t=\tau$) expansion of the corresponding MSD \cite{Balubook}.
We find, in terms of $\tau$, a rather universal value for all the systems 
except for Si, whose more open structure leads to higher Einstein frequencies. 
Therefore, the increase in the MRT of Al and Tl is related to the increased 
barrier height, whereas in Si the increased barrier height is counterbalanced 
by an increased Einstein frequency, leading to MRT similar to those of 
alkalis and alkaline-earths.

The absence of well defined layers in the LJ profile suggests that the motion perpendicular to the interface is much easier than in metals, as no barriers are present, and therefore the position of our arbitrarily defined ``layers" 
has very little influence on the MRT.

\begin{figure}
\begin{center}
\mbox{\psfig{file=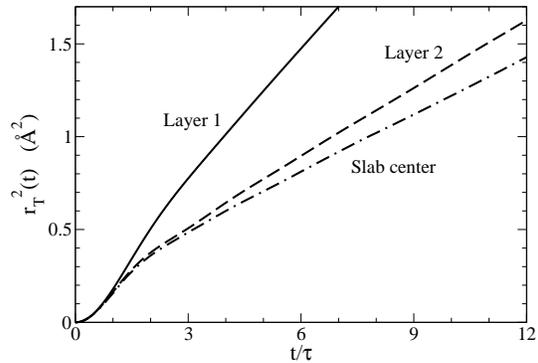,angle=-90,width=85mm}}
\end{center}
\caption{Mean square displacements for layers 1 and 2 and the slab center for
liquid K.} 
\label{r2t}
\end{figure}

Similar trends are also obtained for the liquid Na$_3$K$_7$ alloy, although 
surface segregation leads to an outer layer of almost pure K so that
Na-related quantities are really irrelevant. 
Indeed the outermost Na layer comprises regions 1 and 2, and therefore the 
numbers shown in table \ref{tablilla} for Na in the 
alloy span columns 1 and 2 together.

The integral $P(t)=\int_t^{\infty} p(u)du$, gives the 
probability that having entered the layer, a particle remains there 
longer than $t$, and is used to obtain $t_{20}$ (see the inset of 
figure \ref{probab}).    
Values for $t_{20}$ shown in table \ref{tablilla} correlate rather well 
with the MRT.

\begin{widetext}

\begin{table}
\begin{tabular}{|c|c|c|c|c|c|c|c|c|c|c|c|c|c|}
\hline 
 \multicolumn{2}{|c}{~} & \multicolumn{3}{|c}{$t_{\rm av}/\tau$} & 
 \multicolumn{3}{|c}{$t_{20}/\tau$} 
 & \multicolumn{3}{|c}{$D_T/D_{\rm center}$} 
 & \multicolumn{3}{|c|}{$\Omega_z(\Omega_T)\tau$}
 \\
 System & $\tau$ &Layer 1 & Layer 2& Layer 3
& Layer 1& Layer 2 &Layer 3
& Layer 1& Layer 2 &Layer 3
& Layer 1 & Layer 2 & Center \\ 
\hline  LJ & 0.167 & 11.1 & 10.6 & 10.5 &  
18.7 & 17.7 & 17.2 & 
1.94 & 1.13 & 0.96 &
---(---) & ---(---) & --- \\
\hline  Li & 0.038 & 24.2 & 14.2 & 13.2 &  
32.4 & 19.7 & 17.4 & 
2.01 & 1.19 & 1.09 &
1.27(1.18) & 1.58(1.51) & 1.56 \\ 
\hline  Na & 0.105 & 17.6  & 11.0 & 10.1 & 
26.3 & 16.6 & 14.6 & 
2.16 & 1.18 &  1.12 &
1.37(1.29) & 1.65(1.59) & 1.60\\ 
\hline  K & 0.170 & 19.5 & 11.6 & 10.8 &  
26.8 & 16.8 & 15.0 & 
2.19 & 1.21 & 1.08 &
1.35(1.28) & 1.64(1.57) & 1.59 \\ 
\hline  Rb & 0.284 & 19.5 & 11.3 & 10.6 &  
27.7 & 16.9 & 15.6 & 
2.10 & 1.22 & 1.08 &
1.36(1.29) & 1.67(1.59) & 1.61\\ 
\hline  Cs & 0.352 & 20.3 & 12.0 & 10.6 & 
32.2 & 17.0 & 15.3 & 
1.81 & 1.13 & 1.07 &
1.41(1.35) & 1.73(1.65) & 1.67\\ 
\hline  Mg & 0.056 & 22.5 & 12.0 & 11.1 & 
33.7 & 17.7 & 15.2 & 
2.27 & 1.29 & 1.09 &
1.33(1.33) & 1.65(1.58) & 1.61\\ 
\hline  Ba & 0.158 & 24.6 & 13.5 & 12.8 & 
25.1 & 16.5 & 14.5 & 
2.02 & 1.22 & 1.08 &
1.30(1.28) & 1.63(1.56) & 1.55\\ 
\hline  Al & 0.045 & 37.3 & 17.6 & 14.4 & 
68.0 & 25.6 & 19.3 & 
2.04 & 1.40 & 1.33 &
1.32(1.46) & 1.66(1.62) & 1.65\\ 
\hline  Tl & 0.187 & 64.7 & 26.1 & 18.7 & 
108.2 & 39.4 & 27.0 & 
1.24 & 1.00 & 1.00 &
1.32(1.46) & 1.65(1.60) & 1.62\\ 
\hline  Si & 0.058 & 21.4 & 9.5 & 7.8 &  
37.4 & 16.3 & 13.3 & 
1.24 & 1.00 & 1.00 &
1.68(1.87) & 2.05(1.94) & 1.97\\ 
\hline  Na@Na$_3$K$_7$ & 0.126 &
\multicolumn{2}{c|}{18.7}  
& 10.5 &  
\multicolumn{2}{c|}{17.4}
& 14.7 & 
\multicolumn{2}{c|}{1.28}
& 1.05 & 
\multicolumn{2}{c|}{1.61(1.57)}
& 1.61\\ 
\hline  K@Na$_3$K$_7$ & 0.176 & 18.9 & 10.2 & 9.8 &  
29.4 & 15.2 & 14.4 & 
2.24 & 1.25 & 1.06 &
1.39(1.28) & 1.71(1.64) & 1.67\\ 
\hline 
\end{tabular} 
\caption{Reference time, $\tau$, mean residence time, $t_{\rm av}$, twenty 
percent
time (see text), $t_{20}$, ratio between $D_T$ and the diffusion coefficient
in the center of the slab, $D_{\rm center}$, and Einstein frequencies
for the systems considered and the different regions. The units of $\tau$ for the LJ system are standard reduced units, and picoseconds for all the other systems.}
\label{tablilla}
\end{table}

\end{widetext}

\subsection{Motion parallel to the interface.}

The $t_{20}$ times have been used to analyse the atomic motion within the layers. 
The MSD have been evaluated up to $0.8 \times t_{20}$, in order to allow 
an adequate averaging over the time origins.
Figure \ref{r2t} depicts the MSD for layers 1 and 2 and 
for the slab center in liquid K. 
There is a clear increase in the slope as the interface is approached. This
is common to all the systems, including the LJ one, and is quantified 
in table \ref{tablilla} which shows the ratio 
between the diffusion coefficient parallel to the interface, $D_T$, 
in the different layers and the bulk diffusion coefficient in the slab 
center (where diffusion is isotropic).

We attribute this $\approx$ 100 \% increase in the diffusion at the 
outermost layer to the reduced coordination of the atoms in the interface.
In all the pure metals, except Si, the number of 
nearest neighbors is $\approx$ 12 at the center of the slab and reduces to
$\approx$ 8 at the outer layer. In the LJ system the cordination number again decreases from $\approx$ 12 to $\approx$ 9. For Si the coordination number is 
$\approx$ 6 at the center and $\approx$ $4.5$ at the outer layer. 
This smaller decrease is reflected in a smaller enhancement of the 
diffusion coefficient at the interface of only 24 \%.
A similar explanation also holds for the Na$_3$K$_7$ alloy. The 
total number of neighbors around a K (Na) atom is roughly constant at 
$\approx$ 12.5 (10.0) up to regions 1, where it sharply decreases. 
Consequently, $D_T$ for K increases rapidly in the outermost layer, 
whereas the value for Na in its outermost layer (regions 1 and 2) does 
not change significantly because very few Na atoms in regions 1 are 
affected by this decrease of coordination.

The case of Tl deserves special attention because the number of neighbors 
decreases from 12 to 8 but $D_T$ increases by only 24 \% in the outermost 
layer. We attribute this modest increase in $D_T$ in this strongly layered 
system to a large influence of the second layer on the atoms of the 
outermost one, an effect similar to that exerted by a wall on a liquid which leads
to strong layering and reduced diffusion.

\section{Conclussions.}

In summary, we have analyzed through ab initio simulations the atomic motion in
the liquid-vapor interfaces of several metals and compared it with that of a LJ system in a similar thermodynamic state. Although the layered
structure is similar to other systems such as liquids on a solid or in 
confined geometries, the dynamic behavior within the layers 
is much more similar to the 
liquid-vapor interface of dielectrics, like the LJ system or water, showing 
an enhanced diffusion in the parallel direction at the interface, which is 
attributed to the reduced coordination of an atom
which favours the transverse movement. In the perpendicular direction, the 
layers are too thin to regard the motion as atomic diffusion, and 
instead the MRT associated to each layer has been determined.
The dynamic behavior of metallic systems along this direction is far different from 
that of LJ systems. The value of the MRT for metals is clearly larger at the
outermost layer, contrary to the case of the LJ system, and increases 
significantly for the polyvalent metals with closed structures.
Finally, the reference time $\tau$ is found to be an excellent time unit, since 
it reveals universal values in several dynamic properties of different systems. 

We acknowledge the financial support of the DGICYT (MAT2005-03415) and the EU FEDER program.


\end{document}